\documentclass[draft,jgrga]{agutex}

\usepackage{amssymb,amsmath} 

  \usepackage[dvips]{graphicx}
  \setkeys{Gin}{draft=false}
\authorrunninghead{DOUGLAS ET AL.}

\titlerunninghead{RAIDS 83.4 nm OII Profiles} 

\begin{document}

%
%

\title{Evaluation of Ionospheric Densities Using Coincident OII 83.4 nm Airglow and The Millstone Hill Radar}
%
%



 \authors{E. S. Douglas \altaffilmark{1}, S. M. Smith \altaffilmark{1}, A. W. Stephan\altaffilmark{2}, L. Cashman \altaffilmark{1},  R. L. Bishop\altaffilmark{3}, S. A. Budzien\altaffilmark{2}, A. B. Christensen\altaffilmark{3}, and J. H. Hecht\altaffilmark{3}, S. Chakrabarti \altaffilmark{1}}

\altaffiltext{1}{Center for Space Physics, Boston University, Boston,  MA, USA.}
\altaffiltext{2}{Space Science Division, Naval Research Laboratory, Washington, D. C., USA.}
\altaffiltext{3}{The Aerospace Corporation, Los Angeles, CA, USA.}

%
%


\begin{abstract}
We test the utility of the OII 83.4 nm emission feature as a measure of ionospheric parameters.  Observed with the Remote Atmospheric and Ionospheric Detection System (RAIDS)  {Extreme Ultraviolet Spectrograph} on the International Space Station (ISS), limb profiles of 83.4 nm emissions are compared to predicted dayglow emission profiles from a theoretical model incorporating ground-based electron density profiles measured by the Millstone Hill radar and parameterized by a best-fit Chapman$-\alpha$ function. Observations and models are compared for periods of conjunction between Millstone Hill and the RAIDS fields-of-view. These RAIDS observations show distinct differences in topside morphology between two days, 15 January and 10 March 2010, closely matching the  {forward} model morphology and demonstrating that 83.4 nm emission is sensitive to changes in the ionospheric density profile from the 340 km altitude of the ISS during solar minimum. We find no significant difference between 83.4 nm emission profiles modeled assuming a constant scale height Chapman-$\alpha$ best-fit to the ISR measurements and those assuming varying scale height.  \end{abstract}

%
%

%

\begin{article}

%
%

\section{Introduction}
Time variations in the density and vertical morphology of the ionospheric
plasma embedded within the Earth's thermosphere have considerable
effect on radio communication, radio
astronomy and space weather forecasting \citep{Lawrence1964, afraimovich2008, belehaki2009}. Profiles of ion and electron densities and temperatures for the entire ionosphere are well resolved by the incoherent scatter radar (ISR) technique \citep{evans1969}. The electron density profile below the ionospheric peak altitude ($h_m$), the bottomside, is directly measurable by ionosondes \citep{breit1926,bibl1978}, and radio sounding from orbit can be used to derive topside parameters  \citep{reinisch2001}.  These techniques face several obstacles. ISR and ionosonde profiling from the ground are limited by the geographic distribution of facilities and their associated fields-of-view. Additionally, while more globally distributed, ionosondes can not measure densities beyond the peak, because they measure the range to the nearest region of a given plasma frequency. Thus, determination of the entire ionospheric profile via ionosonde requires the assumption of topside ionospheric parameters or a coincident radio sounding from orbit.  A variety of ionospheric models exist, which depend on  solar and terrestrial inputs and chemistry (see the review by \cite{belehaki2009}). Modeling also faces several challenges, including a lack of topside scale height accuracy in certain cases, as in the presence of strong neutral winds \citep{mikhailov2000}, or validity limited to certain  {geographic} locations for empirically derived models, e.g. \cite{zhang2005}.  Additional measures of the  state of the ionosphere are needed to generate a truly global picture of the space environment.

Measurement of the singly ionized oxygen (O$^+$) emission feature, OII 83.4 nm, has the potential to allow global monitoring of ionospheric parameters in regions where the upper atmosphere is sunlit. This emission is among the brightest features in the terrestrial extreme ultraviolet (EUV) regime between 10 nm and 100 nm. Energetic solar photons  ($\lambda < $45 nm) photoionize inner shell electrons of atomic oxygen via O$(^3P)+h\nu\rightarrow$O$^+(^4P)+e$ \citep{dalgarno1964}, a process that peaks in the 150-175 km altitude range \citep{kumar1983,anderson1985}.  This  leads to an 83.4 nm photon from the $2s^1 2p^4 \ ^4 P \rightarrow  2s^2 2p^3 \ ^4S$ transition, an allowed triplet. The emitted photon subsequently  undergoes resonant scattering by ground-state O$^+$, the dominant ion in the F2 region of the ionosphere with a density profile that closely matches that of electrons.  The peak density in the F2 region typically occurs at higher altitudes (200-500 km),  where the effective recombination coefficient is lower due to diffusion \citep{yonezawa1959}, setting up a separation between the photon production region and scattering region that allows these EUV photons to effectively illuminate the F2 region from below. Since the optical depth, $\tau$, is on the order of 1-10 [Meier, 1991],  scattering leads to an observable altitude profile of 83.4 nm airglow that depends on the O$^+$ ion density, and thus it is expected the distribution of O$^+$ is retrievable  via inversion of a measured airglow altitude profile (e.g. \cite{mccoy1985}). 

Given sufficient constraints, a  unique O$^+$ density profile may be retrieved via radiative transfer analysis \citep{vickers96,picone1997,picone2008,stephan2011a}. For example, as discussed in \cite{stephan2011a} constraining the initial source intensity, due to photoionization, should allow inversion of an 83.4 nm emission profile to return a  unique plasma density profile.
The Remote Atmospheric and Ionospheric Detection System (RAIDS) EUV Spectrograph provides continuous observations of the limb profile of 83.4 nm dayglow, greatly expanding on the previous observations of this emission by sounding rocket \citep{cleary1989, dymond2000, dymond2001, yamazaki2002} and satellite \citep{kumar1983,mccoy1985}.

The purpose of this study is to demonstrate the sensitivity of 83.4 nm emission to ionospheric parameters by comparison to simultaneous ISR measurements as ``ground truth''.   EUV Spectrograph altitude profiles from 15 January 2010 and 10 March 2010 are compared to a model derived from both geophysical parameters and ionospheric plasma density measured at the mid-latitude Millstone Hill Incoherent Scatter Radar (42.6 N geodetic latitude, 288.5 E geodetic longitude; 54~$\Lambda$;  L~$\sim$~3.5). Since densities of the more massive molecular species, such as $N_2$,  become significant at altitudes below the $F_2$ region peak \citep{donahue1968}, the validity of using electron density as a proxy for O$^+$ density is addressed. Both RAIDS and ISR observations are described in Section 2. Fitting the ISR density measurements to a Chapman-$\alpha$ profile and the resulting emission model analysis is described in Section 3. Results of modeling and comparison to the RAIDS EUV observation are presented in Section 4. The implication of these results for ionospheric remote sensing is discussed in Section 5. Findings are summarized in Section 6.

\section{Observations}
	 RAIDS is a suite of instruments spanning the EUV to the near-infrared mounted on the Japanese Experiment Module (\emph{Kibo}) - Exposed Facility (JEM-EF) on the International Space Station (ISS). RAIDS views the anti-ram direction, observing the Earth's trailing limb. One of the instruments, the EUV Spectrograph, covers 55-115 nm and began observations in late-October 2009.  The instrument has a field-of-view of 0.1$^{\circ}$ (altitude) by 2.3$^{\circ}$ (azimuth) and a stepper motor vertically scans  across 16.5$^{\circ}$ in altitude, corresponding to observation of tangent point altitudes from approximately 75 km to 325 km.         
         A description of the RAIDS EUV spectrograph is given by \cite{christensen_instrumentation_1992}, and \cite{budzien2009} detail the RAIDS suite and its ISS deployment.  {The pre-launch instrument characterization showed a  {responsivity} of 0.54$\pm$0.11 counts/sec/Rayleigh at 83.4 nm and a spectral resolution of 1.2 nm \citep{stephan2009}.} This resolution is insufficient to resolve the individual components of the 83.4 nm triplet, 83.276 nm, 83.333 nm, and 83.446 nm \citep{meier1991}.   {Analysis early in the mission showed a time-averaged 0.20\%/day degradation rate of the sensor responsivity.   Dedicated  {responsivity} tests in early 2011 found that the rate of change was dependent on the measured photon flux through the sensor, suggesting a pixel-dependent effect caused by gain changes on the microchannel plate detector and the valid-count pulse height filter implemented in the onboard processing  \citep{stephan2011b}.}  Thus, the degradation rate of a bright feature on the detector, such as the OII 83.4 nm line, is expected to differ from that of neighboring regions on the detector. We found the initial  {responsivity} near turn on, 27 October 2009, and degradation rate by applying a linear regression to each pixel's  {responsivity}  {to ionospheric emissions from comparable look directions under similar solar conditions. This method and the days of comparison are the same as those used by  \cite{stephan2011b} to find the integrated detector  {responsivity} degradation}. This  {responsivity} change is multiplied by the ground based calibration, enabling conversion from instrument counts to Rayleighs for each spectral component on any given day.
                   
\begin{figure}
 \noindent\includegraphics[width=30pc]{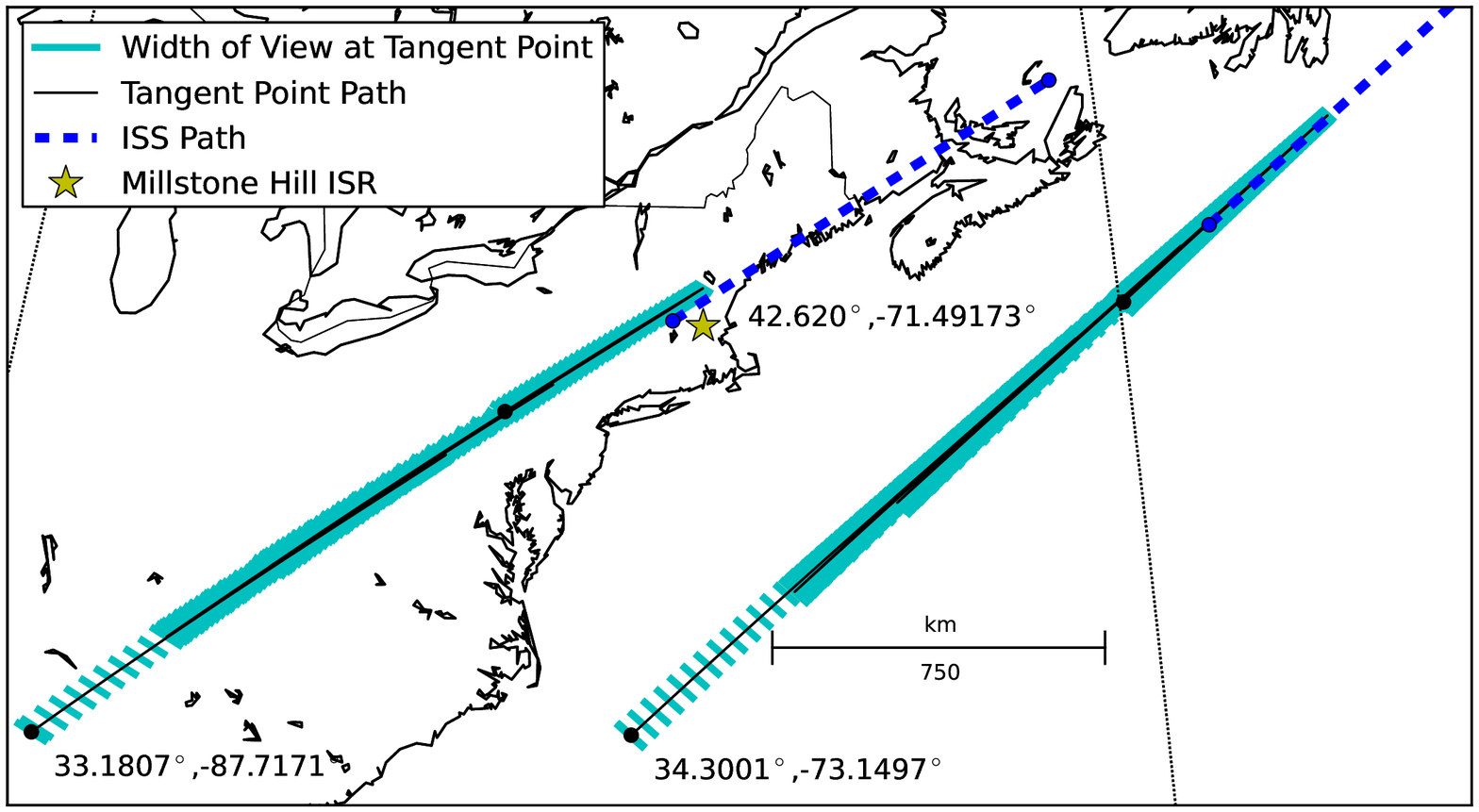}
 \caption{The paths of the observed tangent point and the ISS orbit on 10 March (Left) and 15 January (Right) are indicated by dashed lines. The latitude and longitude of the furthest measurements and of the Millstone Hill Observatory Incoherent Scatter Radar are specified.}
  \label{fig:map}
 \end{figure}

         	The EUV observations analyzed here were collected on 15 January 2010 from 18:59:37.0 UT to 19:02:29.0 UT and 10 March 2010 from 21:42:9.0 UT to 21:44:32.0 UT. The range of observed tangent point Solar Zenith Angles (SZAs) in each period spans from 60 to 90 degrees. Details for each observation, including the position of the tangent point and geophysical parameters are shown in Table 1.  During these RAIDS overflights,  the line-of-sight tangent point passed within 500 km of the  Millstone Hill ISR (i.e. within $\sim$2.5 degree latitude or longitude), allowing for approximate comparisons of the observed  {column emission rate} profiles to the expected  {column emission rate} from ``ground truth" measurements via modeling. F10.7 values were also comparable for these solar minimum times. This similarity in geophysical conditions leads one to expect that the observed EUV profile will be driven by changes in the $F_2$ region profile. The track {s} of the ISS and the RAIDS EUV line-of-sight tangent  point  {are shown} with respect to MIT Millstone Hill Observatory (Local Time = UT $ -5$ hours) in Figure \ref{fig:map}.  
        
       For the study intervals, Millstone Hill was operating a standard zenith profiling experiment using interleaved alternating and long pulse codes designed to monitor the E and F regions of the ionosphere. The alternating pulse is used in this analysis for its higher resolution in the lower ionosphere, where the signal-to-noise ratio is high. The alternating code measurements have characteristic time resolution of 4 minutes and altitude resolution of 4.5 to 58.5 km. Electron density, range (altitude), and associated statistical uncertainty from measurements processed by Millstone Hill's standard INSCAL ISR analysis program were obtained from the Madrigal database \citep{holt2006}. There is an additional uncertainty in measured $N_e$ of up to 10\% arising from uncertainty in the ISR's calibration constant, which was not included in  {the} calculations or figures. 
      
\begin{figure}
 \noindent\includegraphics[width=20pc]{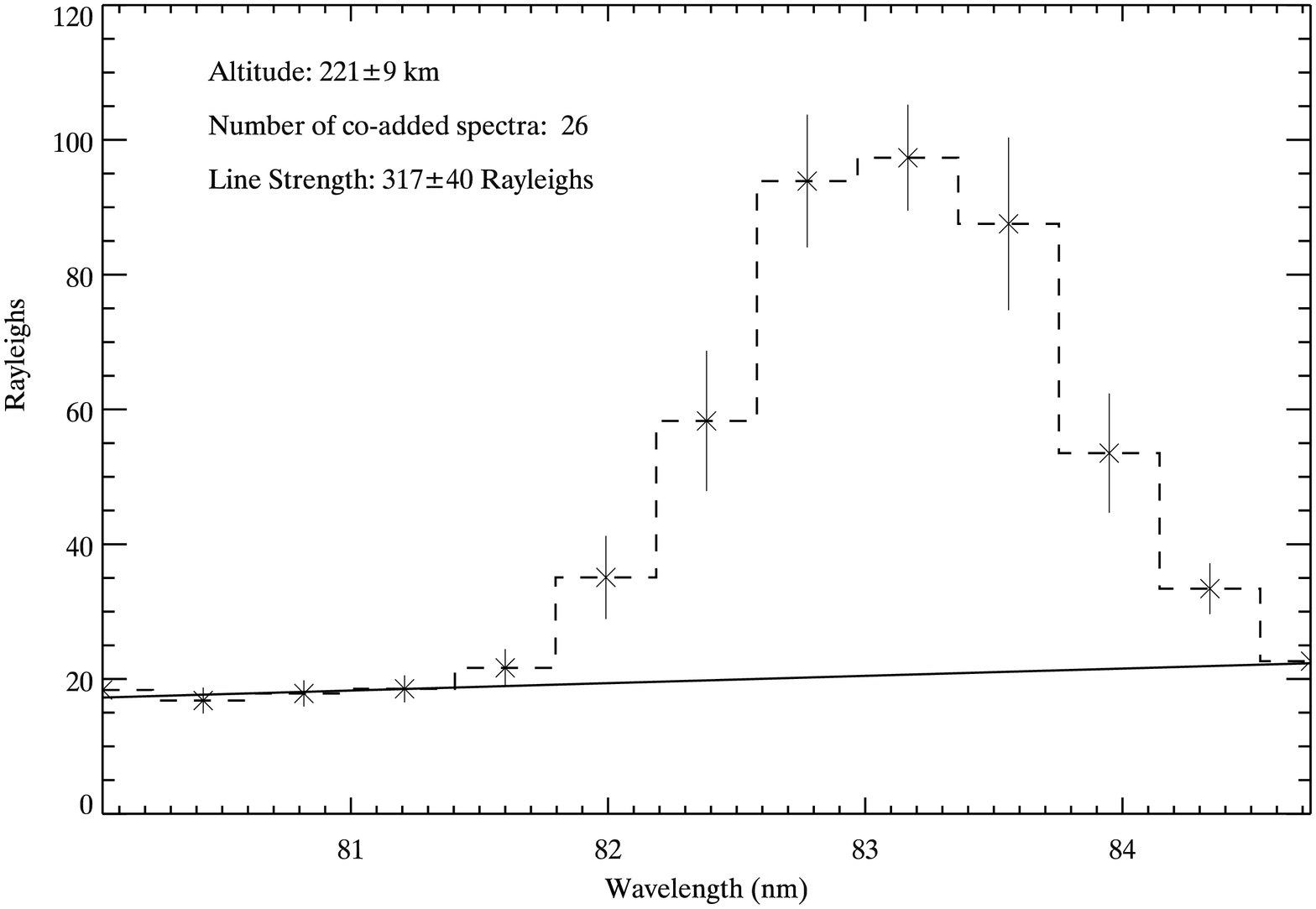}
 \caption{{Sample} mean spectrum of observations within a 20 km altitude range. Calculated {emission in Rayleighs} in each EUV pixel is shown with combined $\pm 1 \sigma$  uncertainty in observation and calibration. {The subtracted background level is the solid line.}}
  \label{fig:spectra}
 \end{figure}

\begin{figure}
 \noindent\includegraphics[width=35pc]{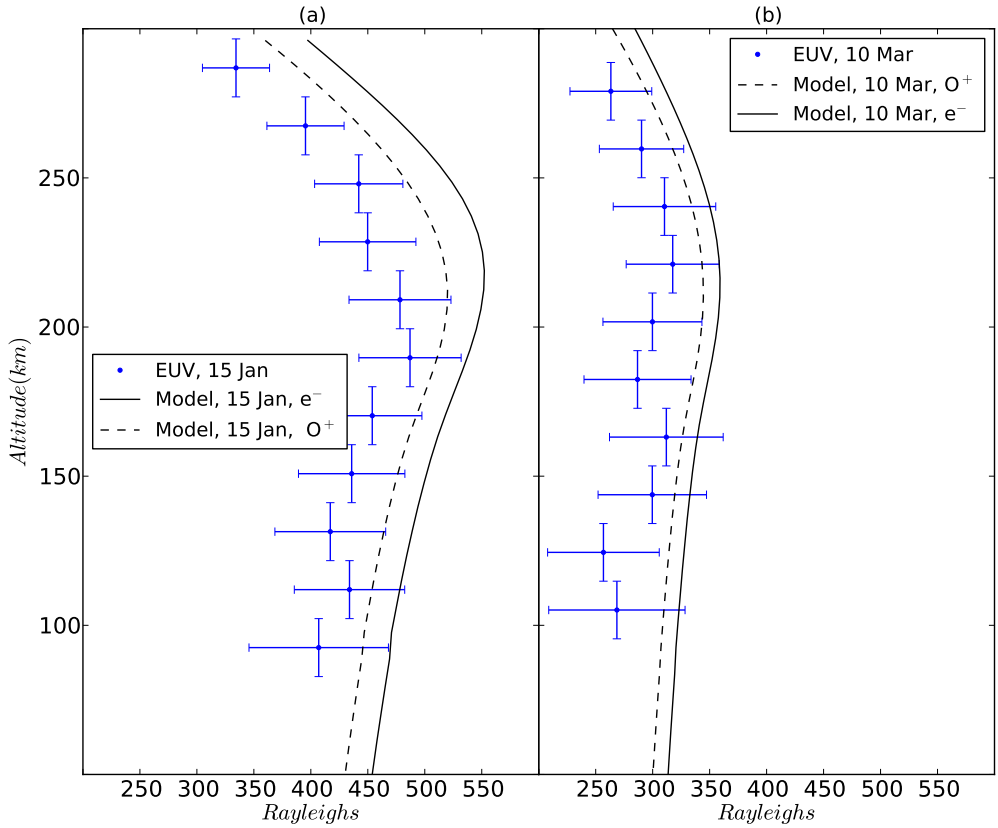}
 \caption{The observed  {and modeled 83.4 nm column emission rate} profiles.  
 { Error bars show 20 km altitude bins and} the $\pm 1 \sigma$ count statistics, {degradation} and calibration uncertainty in the EUV line measurements within each altitude bin. The solid line represents the electron density derived {column emission rate profile}, using the three-parameter Chapman-$\alpha$ ISR profiles as inputs. The dashed line represents the modeled emission profile adjusted to account for the expected differences between O$^+$ and electron density.}
  \label{fig:2panel}
 \end{figure}

	RAIDS EUV spectra within an 80.0 nm to 84.8 nm window were averaged into 20 km tangent point altitude bins. Instrument counts in each pixel of the spectrograph were converted to Rayleighs by applying the degraded calibration. 
	 A linear background was fit to the baseline of each altitude's mean spectrum.
	 An example averaged spectrum and baseline are shown in Figure \ref{fig:spectra}.  The baseline-subtracted signal within the window was summed to derive the total line  {column emission rate} in Rayleighs. Error bars represent $\pm 1 \sigma$ Poisson counting uncertainty, including the uncertainty in the degradation rate of each bin. The discrete points in Figure \ref{fig:2panel} show the calibrated  {column emission rate} profiles, with error bars representing the uncertainty in the observed emissions and the extent of vertical altitude bins.
        
\section{Analysis}
\begin{figure}
 \noindent\includegraphics[width=40pc]{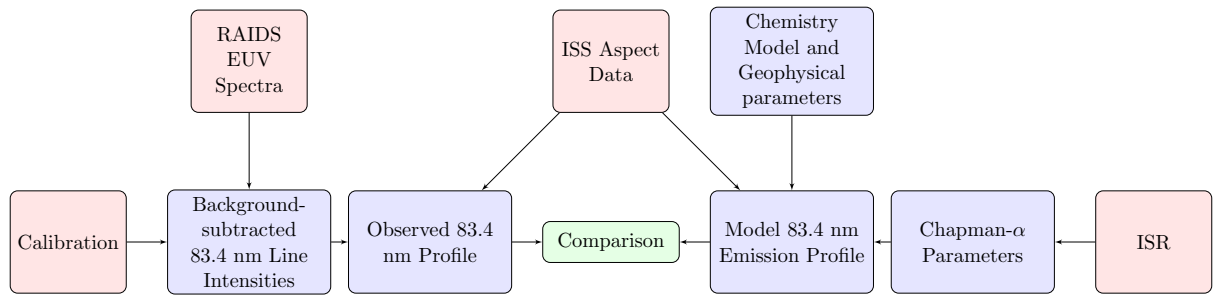}
 \caption{Flow chart of analysis procedure.}
  \label{fig:flow}
 \end{figure}

	  The sequence of analysis which leads to the comparison of the ISR driven emission model with the RAIDS observation is summarized as a flow chart in Figure \ref{fig:flow}.
	 	 In order to compare expected emission to the observed RAIDS emission profiles,  a Chapman-$\alpha$ function was fit to ISR electron density measurements. ISR data   {was selected to span from the lowest observed tangent point to above the ISS orbit, altitudes ($h$) from 100 km to  390 km}. While lacking  {physical} processes such as diffusion, the analytical Chapman function provides a straightforward analytic representation of the ionospheric density profile. The simplest form of this function consists of three parameters, peak density ($N_m$), scale height ($H$) and peak height ($h_m$) as described by \cite{risbeth1969}. The Chapman-$\alpha$ function has the form:
	
	 \begin{equation}
N_e(z)=N_m e^{(\frac{1-z-e^{-z}}{2})}
 \end{equation}
	Where	 \begin{equation} z(h)=\frac{h-h_m}{H}  \end{equation}

	 This canonical function with three constant parameters is preferred for inversion because it provides a sufficiently constrained function as discussed by \cite{stephan2011a}. However, a five-parameter fit of a Chapman-$\alpha$ model with linearly varying scale heights on the top ($A_{1}$) and bottom  ($A_{2}$) of the $F_2$ profile provides a better fit to the ISR measurement, as it accounts for variations in diffusion rate and temperature with height \citep{fox1994}. This function provides an excellent approximation of the variation in the constituent scale height and is tested and discussed extensively by \cite{lei2004}.  The constant scale height $H$ is replaced by the variable $H(h)$, where
	 
\begin{equation} \label{eq:linear_eqs} 
	  H(h)=
	  \begin{cases} 
	  A_1(h-h_m)+H_m & \text{if } h > h_m \\
	  A_2(h-h_m)+H_m & \text{if } h < h_m \\
	  \end{cases}
 \end{equation}

	 \cite{anderson1987} proposed a similar model with six free parameters; however, \cite{fox1994} found better results with Equation \ref{eq:linear_eqs}, which constrain {s} the scale height to equal $H_m$ at $h_m$. Since the goal was an accurate recovery of ionospheric density, the scale height, peak height and peak density were all left as free parameters. These two analytic functions were fit to ISR electron density measurements with statistical errors provided by the Madrigal database from the INSCAL incoherent scatter autocorrelation function analysis program. The fit was optimized using the Levenberg-Marquardt algorithm to minimize $\chi^2$  in the Sherpa modeling and fitting application \citep{refsdal11} in Python. Figures \ref{fig:6panel}(a)  and \ref{fig:6panel}(d)  compare the radar profile and the two Chapman-$\alpha$ models. The fits were {performed} on ISR electron density measurements {in 10 km bins} using the $\pm1\sigma$ uncertainty in the electron density (represented by horizontal error bars). The range resolution of the ISR was not included in the fitting but is represented by vertical error bars for completeness. {The five-parameter fit clearly exhibits better matching of the profile (reduced-$\chi^2=$ 6.9 and 7.7 on 15 January and 10 March, respectively) on both days, recovering both $h_m$ and $N_m$, whereas the classic three-parameter Chapman$-\alpha$ fit recovers $h_m$ to 10\% but underestimates $N_m$ by nearly 25\% on 10 January 2011 and 35\% on 15 March (reduced-$\chi^2= 34$  and 56 on 15 January and 10 March, respectively).}
	   However, this difference is expected to have a small effect; Picone et al. [1997] have shown that adding bottomside parameters which can vary independently of those on the topside does not significantly impact the 83.4 nm profile. Another question is whether the presence of molecular ions at lower altitudes significantly modifies the scattered radiation field when compared to the simpler assumption of a one-to-one relation between O$^+$ in the $F_2$ region. This was tested by repeating the above analysis with an adjusted plasma density, found by subtracting the expected molecular fraction, from the Madrigal database, e.g. \cite{oliver1975}, from the electron density measurements, giving a remaining density closer to the true O$^+$. 
	   
	Both the three and five-parameter best-fit Chapman$-\alpha$ profiles were used as inputs to {an} 83.4 nm airglow model \citep{picone1997} to calculate expected emission. This model serves as a bridge between the ISR and RAIDS data, allowing the direct comparison of ionospheric measurement via the observed O$^+$ signature.
The 83.4 nm model incorporates the most significant photon production, loss, and scattering processes to compute the expected {column emission rate} of the observed emission.  It is designed for the inversion of 83.4 nm airglow emission profiles to derive ionospheric density profile, but can also be used, as is done for this study, to calculate {emission} for direct comparison to EUV observations. 
The airglow model can use any representation of the O$^+$ number density as a function of altitude.  In the inverse problem this is implemented as an analytic function with a 
 number of input parameters that can be used by the model to modify the ion density profile.  The thermosphere is defined by the empirical NRLMSIS thermosphere model \citep{picone2002}, driven by time, day of year, geographic location, in this case Millstone Hill, and solar (F10.7) and geomagnetic (Ap) indices.  
 
The model first calculates the volume excitation rate of the  O$^+$($^4$P) state by ionization of ground state oxygen in the lower thermosphere that leads to the emission of 83.4 nm photons.  This includes photoionization by solar extreme-ultraviolet radiation that dominates the production, as well as photoelectron impact ionization which is the {main} secondary source.  {Resonant scattering of solar 83.4 nm photons was neglected as it is expected to be as much as a factor of 30 lower at solar minimum for these altitudes than the contribution from photoionization \citep{meier1991, link1994}.   }
The model uses a database that provides coefficients for determining the number of excitations per atom, defined as the g-factor, $g(h)$, for the emission as a function of solar F10.7, solar zenith angle, and total column number density of neutrals above the source location \citep{majeed1993}.  The product of the g-factor and the local oxygen density, $N_{O}$, thus defines the volume excitation rate of the initial source. The excitation rate is specified on a discrete grid of altitudes, $h$, that is used to solve the following integral equation for the volume emission rate, $j_k(h)$ \citep{anderson1985, meier1991}:

\begin{equation}
j_k(h) = j_{0k}(h) +\ \sigma_{0k}N_{O^+}(h) \int_{h_0}^\infty j_k(h')H(| \tau_k'- \tau_k|,| t_k'- t_k|)dh' 	  
 \end{equation} 

where $k$ denotes each line of the 83.4 nm triplet.  In this equation, $h_0$ is the altitude below which the photoionization source is negligible or extinction of 83.4 nm photons is large; $\sigma_{0k}$ is the resonant scattering line-center cross section; $\tau_{k}(h)$ is the vertical resonant scattering line-center optical depth at altitude $h$; $t_k(h)$ is the optical depth for vertical pure absorption by $N_2$, O, and O$_2$; and $H$ is the Holstein probability function \citep{holstein1947}, or transport kernel \citep{strickland1970},  that describes the probability that a photon will propagate from region ($h'$, $h'+dh'$) to region ($h$, $h+dh$) and undergo resonant scattering at that point.  This equation assumes isotropic, conservative scattering and complete frequency redistribution in a plane parallel atmosphere.  The first term on the right-hand side includes the initial photon production term from all sources while the second term accounts for the propagation of photons from other altitudes to a given altitude, including the effect of multiple resonant scattering by  O$^+$.

The column emission rate, $4\pi I$, that would be measured by a sensor at position, {\bf r}, along a look direction, ${\bf \hat{e}}$, is calculated by integrating the resulting volume excitation rate along the line of sight via:

\begin{equation}
4\pi I({\bf r, \hat{e}})= 10^{-6}\int \sum\limits_k j_k[{\bf r'}(s)]T_k({\bf r'},{\bf r})d{\bf s}
 \end{equation}

$T_k$ is the probability that a photon traveling along the line of sight ${\bf \hat{e}}$, from ${\bf r}'$ will arrive at  ${\bf r}$ at the location of the observing instrument.  Since $I$ is the radiance in $megaphoton \ cm^{-2} s^{-1} ster^{-1}$, $4\pi I$ is the modeled {column emission rate} in Rayleighs. This model output is used to generate an altitude emission profile for the observed tangent point.  
{Various studies, (e.g. \cite{kumar1983, meier1991, link1994} and \cite{picone1997}), have shown examples of expected observations for various scattering and absorption optical depths from different viewing geometries using comparable emission models.} 
\section{Results}
          Calibrated observations and modeled emission{s} are shown as altitude profiles in Figure \ref{fig:2panel}. The three-parameter fit constants were fed into the model to compute expected 83.4 nm {column emission rate}, shown as a solid lines.  The statistical uncertainty in the RAIDS data arising from Poisson noise and degradation is propagated through the averaging and background {subtraction}.
           The dashed lines represent the model output using the molecular fraction corrected O$^+$ profiles as an input, while the solid line {represents} a one-to-one electron to O$^+$ density relation and used the ISR electron density directly. On both days, the molecular fraction corrected model is comparable to the electron measurements at the certainty level of the RAIDS data.  On 15 January, the increased difference between the profiles may be attributable to the particularly low $h_m$ ($\approx 200km$), partially embedding the $F_2$ region in the molecular ion layer.   On both days measurements are consistently below both model output emission profiles and the morphology is comparable. Compounding calibration errors make this absolute {column emission rate} comparison difficult. The ISR electron density input has up to 10\% calibration error {and significant range uncertainty. The EUV Spectrograph absolute calibration error is 20\%, and the degradation rate changes non-linearly with time, as it is driven by observing conditions and is negligible during times when the sensor is not in operation} \citep{stephan2011b}. Moreover, the absolute scale of the modeled {column emission rate} profile depends on measurements of solar irradiance, and the total column density of neutrals. 
          
\begin{figure}
 \noindent\includegraphics[width=40pc]{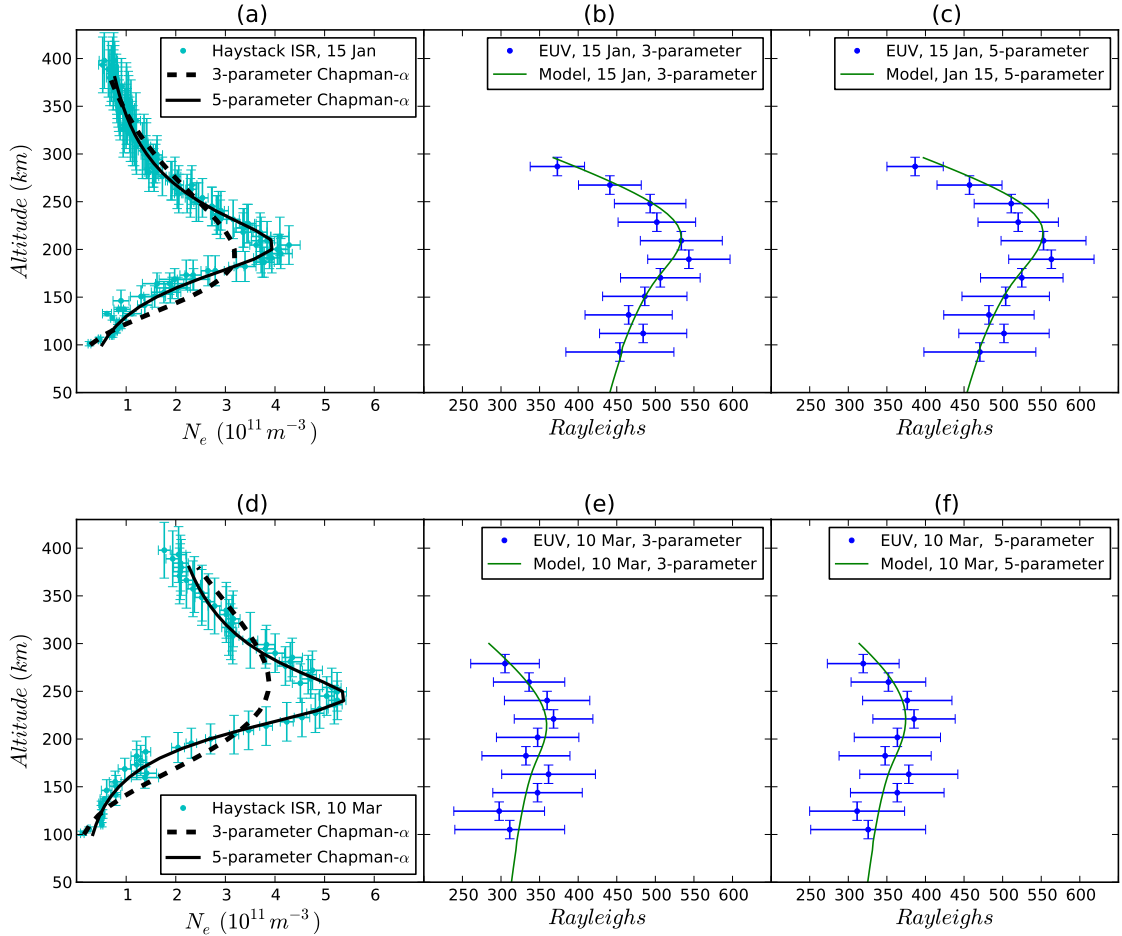}
 \caption{ {\ref{fig:6panel}(a) shows the Millstone Hill ISR electron density profile and Chapman-$\alpha$ fits on 15 January 2010. The three-parameter fits were used as the prescription ionospheric density profile in \ref{fig:6panel}(b), which shows the observed 83.4 nm column emission rate and modeled profiles. The solid line represents modeled emission, using the Chapman-$\alpha$ profiles as inputs.  The five-parameter fit was used as the prescription ionospheric density profile in \ref{fig:6panel}(c).  \ref{fig:6panel}(d-f) are the same as  \ref{fig:6panel}(a-c) but for 10 March 2010.} }
 \label{fig:6panel}
 \end{figure}

           To mitigate these uncertainties in the  {calibrated column emission rate} and allow comparison of model and observations, the {calibrated {column emission rate}} was rescaled with a non-linear $\chi^2$ fit to the modeled emission. These rescaled observations are shown in Figures \ref{fig:6panel} (b, c, e, and f). {The  rescaling factors for the three-parameter fed model, as a ratio of modeled Rayleigh per observed Rayleigh are 1.12$\pm0.07$ and 1.16$\pm0.10$ for 10 March. The five-parameter model rescaling factors, are 1.16$\pm0.07$ and 1.21$\pm0.10$  for 10 March. (These rescaling factors are within the absolution calibration uncertainty.)} 
           On both days the topside profiles agree well and the overall shape is well approximated.  The e-folding  scale height on 15 January is clearly much shorter than on 10 March. Comparison of the two days shows that the observations are morphologically responsive to changes in ionospheric parameters as the observed shape closely tracks the expected emission profile, particularly on the topside. Low altitude emissions vary from one altitude to the next, however, the uncertainty in the {column emission rate} profiles preclude further interpretation of this possible variability. 
           
		  Figure \ref{fig:6panel}(b and e) show modeled emissions using each day's three-parameter Chapman-$\alpha$ fit as input. Figure \ref{fig:6panel}(c and f) shows the result of modeling with five-parameter Chapman-$\alpha$  fits.   Comparison of these profiles show that the three-parameter fit provides sufficient ionospheric information to generate a model which well approximates the 83.4 nm observations in shape. The differences in {column emission rate} between the three and five-parameter fit {models are small compared to the observational $\pm1\sigma$ error bars for these two solar minimum days}.  This is in agreement with the finding of \cite{picone1997}, that a constant ionospheric scale height is sufficient for 83.4 nm modeling. {While providing a useful test of the forward model given ground truth, these two days are poor candidates for inverted retrieval of the ionospheric density from airglow observations, as \cite{picone1997} found that for $h_m$ below 300 km the uncertainty in retrieved parameters prevents a unique inversion.}
		 
\section{Discussion}
The sensitivity of 83.4 nm profiles to ionospheric parameters indicates that it can be a practical means of recovering global ionospheric parameters during the daytime. 
 Future work will serve to compare RAIDS observations to the global ionosonde network and other ISR facilities to further validate {OII 83.4 nm airglow emission} as a global probe of ionospheric parameters. Observations of {OII 83.4 nm airglow emission} by RAIDS will serve to quantify the ionospheric plasma parameters, particularly the more infrequently measured topside profile, which have importance in ionosphere-magnetosphere coupling and satellite communications. As seen in Figure \ref{fig:6panel} and as discussed by \cite{vickers96}, the best-fit three-parameter Chapman-$\alpha$ function accurately captures the expected {emission profile morphology. Additionally, despite the difficulties in calibration, the {column emission rate}  profiles  generated from forward modeling come within 21\% of the calibrated observations, Figure \ref{fig:2panel}.} This approach shows considerable potential, especially considering the 10\% calibration error of the ISR profile and the $\pm$20\% calibration error of the EUV instrument. {By adding free parameters,} inversion to recover a five-parameter model would decrease the number of constraints on the uniqueness of an inverted model. However, the potential exists for constraining the inversion with bottom side sounding.    
 Of great interest is constraining the source intensity of 83.4 nm emission using simultaneous measurement of the altitude profiles of the OII 61.7 nm feature  ($3s ^2 P \rightarrow 2p^3 \ ^2D$), also with the RAIDS EUV spectrograph. This optically thin feature is excited by processes similar to the 83.4 nm feature and thus offers a constraint on the photoionization rate \citep{stephan2011a}.
The accuracy to which the topside was modeled and the insensitivity to low altitude density differences between O$^+$ and electron densities both suggest that the 83.4 nm is most sensitive to high altitude morphology {and integrated plasma density}. This result might be expected since the bulk of photons scattering off O$^+$ ions in the lower altitudes are obscured by absorption along the path to the sensor. Thus, besides {its} global coverage, RAIDS also provides an ideal complement to those ground-based ionospheric sounding methods which are limited to measuring bottomside densities in restricted geographic regions.
 
 \section{Summary}
This comparison of emission profiles from 15 January and 10 March 2010 confirms the utility of the RAIDS EUV instrument and the OII 83.4 nm {airglow} emission line as a measure of key ionospheric parameters. The OII 83.4 nm emission profile observed by RAIDS clearly respond{s} to changes in the ionospheric density profile and its morphology agrees well with model predictions of 83.4 nm emission generated from  Millstone Hill ISR electron density measurements.  


%
%
%
%
%
%
%

\begin{acknowledgments}
The authors would like to thank P. J. Erickson for useful comments, and MIT Haystack Observatory for the use of Millstone Hill ISR data through the Madrigal distributed database system
and its Python API.   Radar observations and analysis at Millstone Hill are supported under Cooperative Agreement ATM-0733510 with the Massachusetts Institute of Technology by the US National Science Foundation under the Geospace Facilities program.
Space Weather Information
from the Space Weather Prediction Center, Boulder, CO, National
Oceanic and Atmospheric Administration (NOAA), US Dept. of
Commerce. Boston University work has been supported by an NSF/UAF
award \#0724440 and the Massachusetts Space Grant Consortium. This work was supported by NRL Base Program work unit 76-9880. Support
for AWS and SAB has been provided by the Office of Naval
Research. RAIDS is integrated and flown as part of the HREP experiment under the direction of the Department of Defense Space Test Program. RAIDS was built jointly by the NRL and The Aerospace Corporation with additional support from the Office of Naval Research.

\end{acknowledgments}

\end{article}


%
%

%

\newpage
 

%
%
%
%
%
%
\begin{table}
\caption{Observation Details}
\centering
\begin{tabular}{l c c}
\hline
 & 15 January 2010 & 10 March 2010  \\
\hline
  Local time & 13:59 to 14:02 UT & 16:42 to 16:44 UT  \\
  Latitude & 34.3$^{\circ}$ N to 45.6$^{\circ}$ N & 33.2$^{\circ}$ N to 43.4$^{\circ}$ N   \\
  Longitude &-73.2$^{\circ}$ to -53.3$^{\circ}$ & -87.7$^{\circ}$ to -71.5$^{\circ}$   \\
Solar Zenith Angle & 62.1$^{\circ}$ to 80.6$^{\circ}$ & 64.1$^{\circ}$ to 79.9$^{\circ}$   \\
F10.7 Flux\tablenotemark{a} & 85 & 80  \\
Ap Index\tablenotemark{a} & 3  &7  \\

\end{tabular}
\tablenotetext{a}{Information from the Space Weather Prediction Center, Boulder, CO,
National Oceanic and Atmospheric Administration (NOAA), US Dept. of Commerce.}
\end{table}



\begin{thebibliography}{}


\bibitem[{\textit{Afraimovich and Yasukevich}(2008)}]{afraimovich2008}
Afraimovich, E. L., and Y. V. Yasukevich (2008), Using GPS-GLONASS-GALILEO data and IRI modeling for ionospheric calibration of radio telescopes and radio interferometers, {\it J. of Atmo. and Solar-Terr. Physics}, 70, 1949-1962, doi:10.1016/j.jastp.2008.05.006

\bibitem[{\textit{Anderson and Meier}(1985)}]{anderson1985}
Anderson Jr., D. E., and R. R. Meier (1985), The OII 834 $\AA$ dayglow: A general model for excitation rate and intensity calculations, {\it Planetary and Space Sci.}, 33(10), 1179-1186, doi:10.1016/0032-0633(85)90075-3.

\bibitem[{\textit{Anderson et al. }(1987)}]{anderson1987}
Anderson, D. N., M. Mendillo, and B. Herniter (1986), A semi-empirical low-latitude ionospheric model, {\it Radio Sci.}, 22(2), 292-306, doi:198710.1029/RS022i002p00292.


\bibitem[{\textit{Belehaki et al.}(2009)}]{belehaki2009}
Belehaki, A., I. Stanislawska, and J. Lilensten (2009), An Overview of Ionosphere-Thermosphere Models Available for Space Weather Purposes, {\it Space Sci. Rev.}, 147, 271-313, doi:10.1007/s11214-009-9510-0.

\bibitem[{\textit{Bibl and Reinisch}(1978)}]{bibl1978}
Bibl, K., and B. W. Reinisch (1978), The universal digital ionosonde, {\it  Radio Sci.}, 13(3), 519-530, doi:197810.1029/RS013i003p00519.\

\bibitem[{\textit{Breit and Tuve}(1926)}]{breit1926}
Breit, G., and M. A. Tuve (1926), A Test of the Existence of the Conducting Layer, {\it Phys. Rev.}, 28, 554-575.


\bibitem[{\textit{Budzien et al.}(2009)}]{budzien2009}
Budzien, S. A., R. L. Bishop, A. W. Stephan, P. R. Straus, A. B. Christensen, and J. H. Hecht (2009), The Remote Atmospheric and Ionospheric Detection System experiment on the ISS: mission overview, {\it Proc. SPIE}, 74380X-74380X-12, San Diego, CA. 

\bibitem[{\textit{Chakrabarti et al.}(1983)}]{chakrabarti1983}
Chakrabarti, S., F. Paresce, S. Bowyer, R. Kimble, S. Kumar, (1983) {The extreme ultraviolet day airglow.} {\it  Geophys Res. Let.}, 88(A6), 4898--4904.



\bibitem[{\textit{Christensen et al. }(1992)}]{christensen_instrumentation_1992}
 Christensen, A. B., D. C. Kayser, J. B. Pranke, P. R. Straus, D. J. Gutierrez, S. Chakrabarti, R. P. McCoy, R. R. Meier, K. D. Wolfram, and J. M. Picone (1992), Instrumentation on the RAIDS (Remote Atmospheric and Ionospheric Detection System) experiment. II - Extreme ultraviolet spectrometer, photometer, and near IR spectrometer. {\it Proc. SPIE},  89-98, San Diego, CA.
 
\bibitem[{\textit{Cleary et al.}(1989)}]{cleary1989}
Cleary, D. D., R. R. Meier, E. P. Gentieu, P. D. Feldman, and A. B. Christensen (1989), An analysis of the effects of N$_2$ absorption on the O(+) 834-A emission from rocket observations, {\it J. of Geophys. Res.},  94, 17281-17285.

\bibitem[{\textit{Dalgarno et al.}(1964)}]{dalgarno1964}
Dalgarno, A., R. J. W. Henry, and A. L. Stewart (1964), The photoionization of atomic oxygen, {\it Planetary and Space Science}, 12(3), 235-246, doi:10.1016/0032-0633(64)90225-9.


\bibitem[{\textit{Donahue}(1968)}]{donahue1968}
Donahue, T. M. (1968), Ionospheric Composition and Reactions,  {\it Science}, 159(3814), 489 -498, doi:10.1126/science.159.3814.489.


\bibitem[{\textit{Dymond et al.}(2000)}]{dymond2000}
Dymond, K. F., R. P. McCoy, S. E. Thonnard, S. A. Budzien, R. J. Thomas, T. N. Bullett, and E. J. Bucsela (n.d.), O$^+$, O, and O$_2$ densities derived from measurements made by the High Resolution Airglow/Aurora Spectrograph (HIRAAS) sounding rocket experiment, {\it J. Geophys. Res.}, 105(A10), PP. 23,025-23,033, doi:200010.1029/1999JA000450.

\bibitem[{\textit{Dymond et al.}(2001)}]{dymond2001}
Dymond, K. F., S. A. Budzien, S. E. Thonnard, A. C. Nicholas, R. P. McCoy, and R. J. Thomas (2001), A Comparison of Electron Density Profiles Derived from the Low Resolution Airglow and Aurora Spectrograph (LORAAS) Ultraviolet Measurements, AGU Fall Meeting Abstracts, 32, 0679.


\bibitem[{\textit{Evans}(1969)}]{evans1969}
Evans, J. V. (1969), Theory and practice of ionosphere study by Thomson scatter radar, {\it Proc. IEEE}, 57(4), 496- 530, doi:10.1109/PROC.1969.7005.

\bibitem[{\textit{Fox}(1994)}]{fox1994}
Fox, M. W. (1994), A simple, convenient formalism for electron density profiles, {\it Radio Sci.}, 29(6), 1473-1491, doi:199410.1029/94RS01666.

\bibitem[{\textit{Holstein}(1947)}]{holstein1947}
Holstein, T. (1947), Imprisonment of resonance radiation in gases, {\it Physical Review}, 72(12), 1212-1233, DOI: 10.1103/PhysRev.72.1212.

\bibitem[{\textit{Holt et al.}(2006)}]{holt2006}
Holt, J. M., L. P. Goncharenko, W. Rideout, and S. Palo (2006), The Madrigal Virtual Observatory - a Fabric for Serving Both Incoherent Scatter and MST Radar Data to the CAWSES Science Community, {\it AGU Fall Meeting Abstracts}, 13, 1165.


\bibitem[{\textit{Kumar et al.}(1983)}]{kumar1983}
Kumar, S., S. Chakrabarti, F. Paresce, and S. Bowyer (1983), The O$^+$ 834-\AA \ Dayglow: Satellite Observations and Interpretation with a Radiation Transfer Model, {\it J. Geophys. Res.}, 88(A11), 9271-9279.

\bibitem[{\textit{Lawrence et al.}(1964)}]{Lawrence1964}
Lawrence, R. S., C. G. Little, and H. J. Chivers (1964), A survey of ionospheric effects upon earth-space radio propagation, {\it Proc. IEEE}, 52(1), 4-27, doi:10.1109/PROC.1964.2737.


\bibitem[{\textit{Lei et al.}(2004)}]{lei2004}
Lei, J., L. Liu, W. Wan, S.-R. hhang, and J. M. Holt (2004), A statistical study of ionospheric profile parameters derived from Millstone Hill incoherent scatter radar measurements, {\it Geophys Res. Let.},  31, 14804, 10.1029/2004GL020578.

\bibitem[{\textit{Link et al.}(1994)}]{link1994}
Link, R., J. S. Evans, and G. R. Gladstone (1994), The O$^+$ 834-\AA \ Dayglow: Revised Cross Sections, {\it J. Geophys. Res.}, 99(A2), 2121-2130, doi:199410.1029/93JA02283.

\bibitem[{\textit{Majeed et al.}(1993)}]{majeed1993}
Majeed, T., D. J. Strickland, J. A. Roberts, R. Link, R. R. Meier, and J. M. Picone (1993), An accurate, fast parameterization of dayglow volume emission rates, {\it Eos Trans. } AGU, 74, Fall Meet. Suppl., 465.

\bibitem[{\textit{McCoy et al.}(1985)}]{mccoy1985}
McCoy, R. P., D. E. Anderson Jr., and S. Chakrabarti (1985), $F_2$ Region Ion Densities from Analysis of O$^+$ 834-\AA \ Airglow: A Parametric Study and Comparisons with Satellite Data, {\it J. Geophys. Res.}, 90(A12), 12257-12264, doi:198510.1029/JA090iA12p12257.

\bibitem[{\textit{Meier}(1991)}]{meier1991}
Meier, R. R. (1991), Ultraviolet spectroscopy and remote sensing of the upper atmosphere, {\it Space Sci. Rev.}, 58, 1-185.

\bibitem[{\textit{Mikhailov et al.}(2000)}]{mikhailov2000}
Mikhailov, A. V., T. Y. Leschinskaya, G. Miro, and V. K. Depuev (2000), A comparison of N$_e$(h) model profiles with ground-based and topside sounder observations, {\it Ann. Geophys}., 43(1),105-117, doi:10.4401/ag-3621. 

\bibitem[{\textit{Oliver}(1975)}]{oliver1975}
W. L. Oliver, Models of F1-region ion composition variations, {\it J. Atmos. Terr. Phys.}, 37, 1065-1076 (1975).

\bibitem[{\textit{Picone et al.}(1997)}]{picone1997}
Picone, J. M., R. R. Meier, O. A. Kelley, K. F. Dymond, R. J. Thomas, D. J. Melendez-Alvira, and R. P. McCoy (1997), Investigation of ionospheric O$^+$ remote sensing using the 834-\AA \ airglow,{ \it J. Geophys. Res.}, 102, 2441-2456, doi:10.1029/96JA03314.

\bibitem[{\textit{Picone et al.}(2002)}]{picone2002}
Picone, J. M., A. E. Hedin, D. P. Drob, and A. C. Aikin (2002), NRLMSISE-00 empirical model of the atmosphere: Statistical comparisons and scientific issues, { \it J. Geophys. Res.}, 107, 16, doi:200210.1029/2002JA009430.

\bibitem[{\textit{Picone}(2008)}]{picone2008}
Picone, J. M. (2008), Influence of systematic error on least squares retrieval of upper atmospheric parameters from the ultraviolet airglow, {\it J. Geophys. Res.}, 113, 17, doi:200810.1029/2007JA012831.

\bibitem[{\textit{Refsdal et al.}(2011)}]{refsdal11}
Refsdal,., R., S. Doe, D. Ng, A. Siemiginowska, D. Burk, J. Evans, and I. Evans (2011), Advanced Python Scripting Using Sherpa, {\it Astronomical Data Analysis Software and Systems XX}. ASP Conference Proceedings, vol. 442, PP. 687, Astronomical Society of the Pacific, San Francisco.

\bibitem[{\textit{Reinisch et al.}(2001)}]{reinisch2001}
Reinisch, B. W., D. M. Haines, R. F. Benson, J. L. Green, G. S. Sales, and W. W. L. Taylor (2001), Radio sounding in space: magnetosphere and topside ionosphere, {\it J. of Atmo. and Solar-Terr. Phys.}, 63, 87-98.

\bibitem[{\textit{Rishbeth and Garriot}(1969)}]{risbeth1969}
Rishbeth, H., and O. K. Garriott (1969), {\it Introduction to Ionospheric Physics}, Academic, San Diego, CA.

\bibitem[{\textit{Stephan et al.}(2009)}]{stephan2009}
Stephan, A. W., S. A. Budzien, R. L. Bishop, P. R. Straus, A. B. Christensen, J. H. Hecht, and Z. Van Epps (2009), The Remote Atmospheric and Ionospheric Detection System on the ISS: sensor performance and space weather applications from the extreme to the near ultraviolet, p. {\it Proc. SPIE}, 74380Y-74380Y-10, 438, doi:10.1117/12.825167, San Diego, CA.

\bibitem[{\textit{Stephan et al.}(2011a)}]{stephan2011a}
Stephan, A. W., J. M. Picone, S. A. Budzien, R. Bishop, A. B. Christensen, and J. H. Hecht (2011), Measurement and application of the OII 61.7 nm airglow,  {\it J. Geophys. Res.}, doi:10.1029/2011JA016897, in press (accepted 15 November 2011).

\bibitem[{\textit{Stephan et al.}(2011b)}]{stephan2011b}
Stephan, A.W. et al. (2011b), Characterization of sensitivity degradation seen from the UV to NIR by RAIDS on the International Space Station. {\it Proc. SPIE},  in preparation.

\bibitem[{\textit{Strickland and Donahue}(1970)}]{strickland1970}
Strickland, D. J., and T. M. Donahue (1970), Excitation and radiative transport of OI 1304 \AA  \ resonance radiation - I. The dayglow, {\it Planet. Space Sci., }18(5) 661-689, DOI: 10.1016/0032-0633(70)90049-8.

\bibitem[{\textit{Vickers}(1996)}]{vickers96}
Vickers, J. (1996), An evaluation of EUV remote sensing of the ionosphere using the O-II
834-\AA \ emission, Ph.D. thesis, University of California, Berkeley.

\bibitem[{\textit{Yamazaki et al.}(2002)}]{yamazaki2002}
A. Yamazaki, S.Tashiro, Y. Nakasaka, I. Yoshikawa, W. Miyake, and M. Nakamura. (2002), Sounding-rocket observation of O {II} 83.4-nm emission over the polar  ionosphere. {\it Geophys. Res. Let}, 29:4.


\bibitem[{\textit{Yonezawa}(1959)}]{yonezawa1959}
Yonezawa, T. (1959), A new theory of formation of the F2-layer, {\it  J. of Atmo. and Solar-Terr. Physics}, 15(1-2), 89-94, doi:10.1016/0021-9169(59)90178-3.

\bibitem[{\textit{Zhang et al.}(2005)}]{zhang2005}
Zhang, S.-R., J. M. Holt, A. P. van Eyken, M. McCready, C. Amory-Mahaudier, S. Fukao, and M. Sulher (2005), Ionospheric local model and climatology from long-term databases of multiple incoherent scatter radars, {\it Geophys Res. Let}, 32, 20102, doi:10.1029/2005GL023603
\newpage
\end{thebibliography}
\end{document}